\documentclass[12pt]{iopart}
\usepackage{graphicx}
\usepackage{subcaption}
\usepackage{epstopdf}
\usepackage{color} %needed just for comments
\usepackage{xcolor}
\usepackage{bm}
\usepackage{cite}
\usepackage{float} %Pour figer les figures avec "H"
\usepackage{enumerate}
\usepackage{pifont}
\usepackage{cuted}
\usepackage{enumitem}
\usepackage[linktocpage=true,hidelinks]{hyperref}

\expandafter\let\csname equation*\endcsname\relax
\expandafter\let\csname endequation*\endcsname\relax
\usepackage{amsmath}
\usepackage{cleveref}

\begin{document}

\title[]{Particle propagation and electron transport in gases}

\author{L. Vialetto$^1$, H. Sugawara$^2$, and S. Longo$^{3,4}$}
\address{$^1$ Aeronautics and Astronautics, Stanford University, 496 Lomita Mall, Stanford, CA 94305, United States of America}
\address{$^2$ Division of Electronics for Informatics, Graduate School of Information Science and Technology,
Hokkaido University, Sapporo 060-0814, Japan}
\address{$^3$ Dipartimento di Chimica, Universit\`a degli Studi di Bari, via Orabona 4, 70126 Bari; Istituto per la Scienza e Tecnologia dei Plasmi del CNR, Via Amendola 122/D, 70125 Bari, Italy}
\address{$^4$ Istituto per la Scienza e Tecnologia dei Plasmi, CNR, 70126 Bari, Italy}

\ead{vialetto@stanford.edu, sugawara@ist.hokudai.ac.jp, savino.longo@uniba.it}
\vspace{10pt}
\begin{indented}
\item[]\today
\end{indented}

\begin{abstract}
In this review, we detail the commonality of mathematical intuitions that underlie three numerical methods used for the quantitative description of electron swarms propagating in a gas under the effect of externally applied electric and/or magnetic fields. 
These methods can be linked to the integral transport equation, following a common thread much better known in the theory of neutron transport than in the theory of electron transport. First, we discuss the exact solution of the electron transport problem using Monte Carlo (MC) simulations. In reality we will progress much further, showing the interpretative role that the diagrams used in quantum theory and quantum field theory can play in the development of MC\@. Then, we present two methods, the Monte Carlo Flux and the Propagator method, which have been developed at this moment. The first one is based on a modified MC method, while the second shows the advantage of explicitly applying the mathematical idea of propagator to the transport problem.

\end{abstract}


\section{Introduction}\label{sec:intro}

The goal of kinetic theory is the modelling of a gas (or plasma, or any system made up of a large number of particles) by a distribution function in the particle phase space. This phase space includes macroscopic variables, i.e.\ the position in physical space, but also microscopic variables, which describe the ``state'' of the particles. In this respect, several textbooks and review papers have been devoted to describe the mathematical foundations of kinetic theory of gases and plasmas \cite{villani2002review, cercignani1969mathematical, desvillettes1996splitting, balescu1963statistical, grad1949kinetic, duderstadt1979transport}. The need to create mathematical models of ionized gases at low temperatures has also been felt for many decades. In fact, the applications of these systems are numerous and qualitatively important and range from the detection of ionizing radiation to the production of new materials, to medicine, and the emission of light \cite{adamovich20222022}. It is not possible to build a reliable model of a low-temperature ionized gas without a detailed description of electron transport and kinetics. For this reason, a gigantic literature was born and developed, including thousands of works, on the development of numerical calculation methods that allow us to calculate the quantities that describe these phenomena \cite{makabe2018velocity, kushner1987application, boyle2023boltzmann}. 

Since the density of electrons in a weakly ionized gas is much lower than that of the neutral background gas, it is usually assumed that they do not collide altogether or with other charged particles. Their behavior is governed by externally applied fields (electric or magnetic) and by the various collisions that they undergo with the molecules or atoms of the neutral gas. 
Under these assumptions, it follows that the motion of electrons in the phase space can be described by a one-particle distribution function $f(\bm{r},\bm{v},t)$ under the linear electron Boltzmann equation (EBE) \cite{braglia1984twoterm, braglia1985multi}:
\begin{equation}
\frac{\partial f(\bm{r},\bm{v}, t)}{\partial t} + \bm{v}(t)\cdot\bm{\nabla}_{\bm{r}}f(\bm{r},\bm{v}, t)
- \frac{e\bm{E}(t)}{m}\cdot\bm{\nabla}_{\bm{v}}f(\bm{r},\bm{v}, t) = J[f],
\label{eq:boltzmann_1}
\end{equation}

\noindent
where $\bm{r}$, $\bm{v}$, and $t$ are the configuration space coordinates, velocity space coordinates, and time, respectively, $e$ is the elementary charge, $m$ is the electron mass, $\bm{E}$ is the electric field, and $J[f]$ is the collision term that takes into account electron collisions with the neutral background gas atoms or molecules.
For the category of problems in question, this equation which is a non-linear equation becomes linear because in the product of distribution functions in the right-hand-side of the equation, one of the two functions describes the neutral gas particles.
Even with this simplification, the equation presents considerable difficulties and this is why the literature relating to its solution is so extensive \cite{segur1986survey, white2009recent, makabe2018velocity, boyle2023boltzmann}.

For many years, and still effectively when appropriate, an approximation known as the two-term approximation has been employed to solve this equation \cite{rockwood1973elastic, hagelaar2005solving, tejero2019lisbon, braglia1984twoterm, colonna2022two}. It consists in simplifying the electron velocity distribution function (EVDF) by developing it into an expansion in spherical harmonics truncated at first order. This approximate methodology is quick and simple to implement numerically and has been available for several years also into open access programs, which allow one to calculate the isotropic and the first anisotropic components of the velocity distribution function \cite{lokiGithub, bolsigWebpage, tejero2019lisbon}. 
Over the years, it has emerged that the two-term approximation is not sufficient in a series of problems especially related to plasmas confined inside reactors, whose geometry must be considered \cite{white2003classical, dujko2011multi, Loffhagen, robson2017fundamentals, pitchford1981extended, stephens2018multi}. For this reason and with the help of the ever-improving performance of computers, more direct calculation methods have been developed. These allow the distribution function of Equation \eqref{eq:boltzmann_1} to be determined if necessary, taking into account all the independent variables involved. 

The first of these methods, still one of the most used today, is the Monte Carlo (MC) method which directly simulates the trajectory of each electron in a large ensemble using random numbers \cite{longo2000monte, yousfi1994monte, boeuf1982monte, penetrante1985monte}. This method is a variation of the one originally developed in the 1940s for calculating neutron trajectories in the simulation of nuclear fission systems \cite{spanier2008monte}. Nowadays, several codes are available as open source for MC simulations of electrons in gases \cite{MAGBOLTZWebpage, biagi1999monte, rabie2016methes, dias2023lisbon}.
Despite the accuracy of MC simulations for calculation of chemical rate coefficients and electron transport parameters and the increase in computational resources, this method is still computationally expensive (especially when coupled with self-consistent description of weakly-ionized gases) \cite{taccogna2015monte}. 

Recently, other deterministic methods for numerical solutions of the EBE have emerged \cite{schaefer1990monte, Sugawara2021, gamba2018galerkin}. For a long time, these methods were essentially unaffordable because of the high computational cost of the integral in the Boltzmann collision operator, but they are now becoming more and more competitive. 
The methods that will be exposed in this review share the role played in their formulation and practice by mathematical entities called Green functions or propagators \cite{challis2003green}. These, basically, are operators that allow to calculate the evolution of a system, as long as it is described by linear equations, by breaking it down into the contributions that come from every small part of the system into every other small part of the system. Green functions, on which there is a gigantic literature, place the methods considered here in the broader perspective of mathematical physics. Furthermore, awareness of the common conceptual basis between these methods produces advantages both for those who develop them, for those who employ them and for those who teach them. In the following sections we will develop this point of view.

To summarize, the numerical methods developed for solving the EBE are an important part of mathematical physics and theoretical physics, and any research in this sector deserves to be considered in the broader context. On the occasion of the 150\textsuperscript{th} anniversary of the original formulation of the EBE, a review article on methods for calculating the kinetics of electrons based on the EBE was published by Boyle and co-authors \cite{boyle2023boltzmann}. 
The present review has been formulated as a complement to that work. In fact, we focus here on methods for numerical solutions of the EBE that do not employ an expansion of the velocity distribution function in spherical harmonics. Moreover, the present review updates the survey written by Segur and co-authors \cite{segur1986survey} with a description of recent efforts in development of numerical methods used to solve the EBE. 

The review is structured as follows. Section \ref{sec:monte_carlo}
describes the principles of MC simulations of electrons and the connection 
of MC methods with the electron transport equation. 
Section \ref{sec:monte_carlo_flux} presents the mathematical principles behind the 
Monte Carlo Flux method, that is a hybrid stochastic-deterministic algorithm to solve the EBE. 
The advantages and disadvantages of this method are also described. 
Finally, in Section \ref{sec:propagator_method}, the Propagator method is described and its 
application for electron swarm and plasmas are highlighted, as well as the numerical method. 
To conclude, we show that awareness of the common conceptual basis between these methods produces advantages both for those who develop them, for those who employ them, and for those who teach them. In the following sections, we will develop this point of view.
 
\section{The Monte Carlo method}\label{sec:monte_carlo}

\subsection{Principle}
The MC method is still today the most used for the description of electron swarms especially in situations of complex geometry or when, in the context of a simulation of a plasma device, electrons are interacting with space charge or with the walls of the device \cite{yousfi1994monte, kushner1987application, boeuf1982monte, longo2006monte, pitchford1982comparative, biagi1999monte, yousfi2009electron, raspopovic1999benchmark, loffhagen2002boltzmann, taccogna2015monte}.

The method is based on the description of the movement of a large number of mathematical objects representing electrons under the effect of forces due to electric and magnetic fields, while collisions with neutral plasma particles are introduced by means of random times between each collisions and the next one.

It is very simple to create a MC simulation \cite{longo2000monte}; it is sufficient to organize a vector of objects representing the electrons or, in a more traditional setting, numerical vectors for the individual dynamic quantities of each electron, i.e.\ $x$, $y$, $z$, $v_x$, $v_y$, $v_z$. Once a small, appropriately chosen time step has been introduced, it is sufficient to take into account the effect of the forces acting on each electron due to its position and speed. At this point, it is sufficient to apply the kinetic theory of gases to determine the probability that a given collision process extracted from the set of cross sections can occur. The effect of this collision process on any single electron is produced explicitly, for example by modifying the velocity components to take into account an elastic or inelastic process.

Alternatively, each electron is associated with the time to its next collision, this is updated during the calculation, and when this is reached the speed of the electron is changed and a new time to the next collision is calculated. the version of the method most often used for precision calculations is the one called null collisions MC. The method was first introduced by Skullerud \cite{skullerud1968stochastic} for the use in plasma, although a similar one, the artificial isotope method \cite{spanier2008monte}, had been previously used for the treatment of complex geometries in nuclear reactors. The basic idea is simple and powerful and consists in introducing a collision process, fictitious, in such a way that its frequency added to that of the other processes produces a collision frequency that does not depend on space, time or speed. This way it is extremely simple to calculate the time to the next process, only once after any collision event, using the equation \cite{longo2002direct}:
\begin{equation}
    t_{\rm c} = - \frac{1}{\nu} \log \eta, 
\label{eq:collision_times}
\end{equation}

\noindent
where $t_{\rm c}$ is the time to next collision, $\nu$ is the above mentioned constant total momentum transfer collision frequency (including the null collision frequency), and $\eta$ is a random number from a uniform distribution in $(0,1]$.
To compensate for the fictitious process that has been added, this last is simply considered equal to the others, but when it occurs, the speed of the electron does not change. This method also allows the effect of the non-zero temperature of the gaseous medium on the electrons to be introduced: it is clear in fact that the collision frequency between a neutral particle and an electron depends on the speed of the particle. This effect is particularly important if ions are propagated instead of electrons \cite{ristivojevic2012monte, longo2004monte}. A reasonable maximum of the collision frequency is then taken, and at the time of the collision a fraction of the events are eliminated to account for the Maxwell distribution of gas particle velocities. 

\subsection{Monte Carlo method as formal solution of the electron transport problem}
The MC method, with the prescription summarized above, is able to solve the problem of the transport of a swarm in an exact way, given that no numerical parameter is introduced that needs to be optimized. For this reason it has sometimes been considered a kind of numerical experiment based on an analogical approach, on the direct use of the laws of physics, something different from the solution of transport equations. This idea is in contrast with the very principles of the initial development of the MC method, at the time of which it was perfectly clear to the developers that the method can be derived by formal procedures starting from the integral form of the transport equation \cite{spanier2008monte}; there is therefore no opposition between an analogical approach and an approach based on the solution of the integral equation, especially since the two methods ultimately lead to the development of the same code. 

In the case of applying the method to swarms of charged particles, there are some formalities to introduce in the mathematical proofs, given that charged particles do not propagate in straight lines unlike photons and neutrons. However, these technical aspects have been addressed in several publications \cite{resibois1977classical, cercignani1969mathematical}. In this regard, an illuminating way of showing how the continuous operators of the EBE become events in the MC method is to use the well-known integral equation \cite{resibois1977classical} satisfied by the time evolution operator $U$ written as the exponential of the sum of two operators $At$ and $Jt$ where $t$ is time:
\begin{equation}
    U(t) = \exp\left(-(A+J)t\right)
    \label{eq:operator}
\end{equation}

\noindent
very well known in quantum field theory, but mathematically true in general;
\begin{equation}
    U(t) = \exp(-At) + \int_{0}^{t}dt' \exp(-A(t-t')) J U(t').
\end{equation}

\noindent
Now if $\exp(-At)$ is the operator which propagates a particle for the time $t$ under the action of inertia and electric force,
while $-J$ is the integral operator representing the collision events, 
\begin{equation}
-Jf = \int_{\bm{v}'} \left[p_{(\bm{v}, \bm{v}'}f(\bm{v}') - p_{\bm{v}', \bm{v}}f(\bm{v})\right] d^3v'.
\label{eq:collision_operator_0}
\end{equation}
\noindent
The equation above shows that the last collision event randomly breaks the propagation into two parts, or stages, one of which, $U(t')$, can still include other collision events in a hierarchical structure.

This structure is resumed in a natural way using a  diagramming technique that was mentioned in the work \cite{longo2000monte} developed in more detail in \cite{longo2008derivation}.

To show this, we introduce, for the first time explicitly, the Green function or propagator between two positions in phase space after a time shift $\Delta t = t' - t$ that is represented here as $G(\bm{r}',\bm{r},\bm{v}',\bm{v},t',t)$ which actually is a generalized function, since it may include the Dirac's delta $\delta(\bm{r}'-\bm{r},\bm{v}'-\bm{v},t'-t)$ in its expression. 

The intuitive aspect of traditional MC methods, derives from the fact that in the absence of collisions the Green function of two arguments final $v',r',t$ and initial $v,r$ is equal to $0$ unless there is an  electron trajectory, determined by inertia and by the electric and possibly magnetic fields, which smoothly joins the two arguments; in which case the Green function has the aforementioned form of the Dirac generalized function. Calculating this propagator without collisions is therefore equivalent to solving the equations of motion for a material point with a charge-to-mass ratio corresponding to an electron, under the effect of fields.

% Here is the intuitive advantage of these direct methods. Effective numerical methods of trajectory tracing can be applied to this calculations. For these statistical calculations it was found by experience that highly accurate methods developed for few-particle calculations, like Runge-Kutta, are often not the best choice; instead, effective velocity vs. accuracy compromises are provided by the second-order Leapfrog method, also very simple to implement and the Boris-Buneman method, which allows to trace exactly the trajectory of a charged particle for finite times under the effect constant and uniform electric and magnetic field, and can be used with an appropriate numerical time step in more general cases.
When collisions of electrons with atoms and molecules are added to the description, the point probability packets are dispersed, which means that the initial Dirac distribution becomes a Dirac distribution multiplied by a value less than 1, added to a traditional function. At this stage, it is no longer possible to use a description based only on deterministic trajectories.%

Now, we can introduce here a notation well known in quantum field theory \cite{berestetskii1982quantum} where $G$, the Green function of the equation, is the exact propagator, while we use $g$ to represent the Green function where the positive part of the collision operator $J$ is removed, so that particles propagate according to the convective operator while they ``decay'' and disappear according to the exponential factor $\exp(-\nu t)$. The result is the expansion:
%
%\begin{equation}\begin{array}
%\\G(\bm{r},\bm{r}_0,\bm{v},\bm{v}_0,t,t_0) = 
%g(\bm{r},\bm{r}_0,\bm{v},\bm{v}_0,t,t_0) +\\
%\int\int\int d^3r_1 d^3v_1 d^3v_2 \int_{t_0}^t dt_1
%g(\bm{r},\bm{r}_1,\bm{v},\bm{v}_2,t,t_1)
%p_{\bm{v}_1 \bm{v}_2 }g(\bm{r}_1,\bm{r}_0,\bm{v}_1,\bm{v}_0,t_1,t_0)+\\
%\frac{1}{2} \int\int\int d^3r_1 d^3v_1 d^3v_2 \int_{t_0}^t dt_1
%\int\int\int d^3r_2 d^3v_3 d^3v_4 \int_{t_0}^t dt_2
%g(\bm{r},\bm{r}_2,\bm{v},\bm{v}_4,t,t_2)\\
%p_{\bm{v}_1 \bm{v}_2 }
%g(\bm{r}_1,\bm{r}_0,\bm{v}_1,\bm{v}_0,t_1,t_0)
%p_{\bm{v}_3 \bm{v}_4 }
%g(\bm{r}_2,\bm{r}_1,\bm{v}_3,\bm{v}_2,t_2,t_1)+
%...\end{array}
%\label{eq:expansion}
%\end{equation}
\begin{align}
& G(\bm{r},\bm{r}_0,\bm{v},\bm{v}_0,t,t_0)
  = g(\bm{r},\bm{r}_0,\bm{v},\bm{v}_0,t,t_0) \notag \\
& + \displaystyle
    \int \! \! \! \int \! \! \! \int
    d^3r_1 d^3v_1 d^3v_2 \int_{t_0}^t dt_1
    g(\bm{r},\bm{r}_1,\bm{v},\bm{v}_2,t,t_1)
    p_{\bm{v}_1 \bm{v}_2}
    g(\bm{r}_1,\bm{r}_0,\bm{v}_1,\bm{v}_0,t_1,t_0) \notag \\
& + \displaystyle
    \frac{1}{2} \int \! \! \! \int \! \! \! \int
    d^3r_1 d^3v_1 d^3v_2 \int_{t_0}^t dt_1
    \int \! \! \! \int \! \! \! \int
    d^3r_2 d^3v_3 d^3v_4 \int_{t_0}^t dt_2
    g(\bm{r},\bm{r}_2,\bm{v},\bm{v}_4,t,t_2) \notag \\
& \times p_{\bm{v}_1 \bm{v}_2}
    g(\bm{r}_1,\bm{r}_0,\bm{v}_1,\bm{v}_0,t_1,t_0)
    p_{\bm{v}_3 \bm{v}_4}
    g(\bm{r}_2,\bm{r}_1,\bm{v}_3,\bm{v}_2,t_2,t_1) + \cdots
\label{eq:expansion}
\end{align}

\noindent

\begin{figure}[h]
    \centering
    \includegraphics[width=0.5\textwidth]{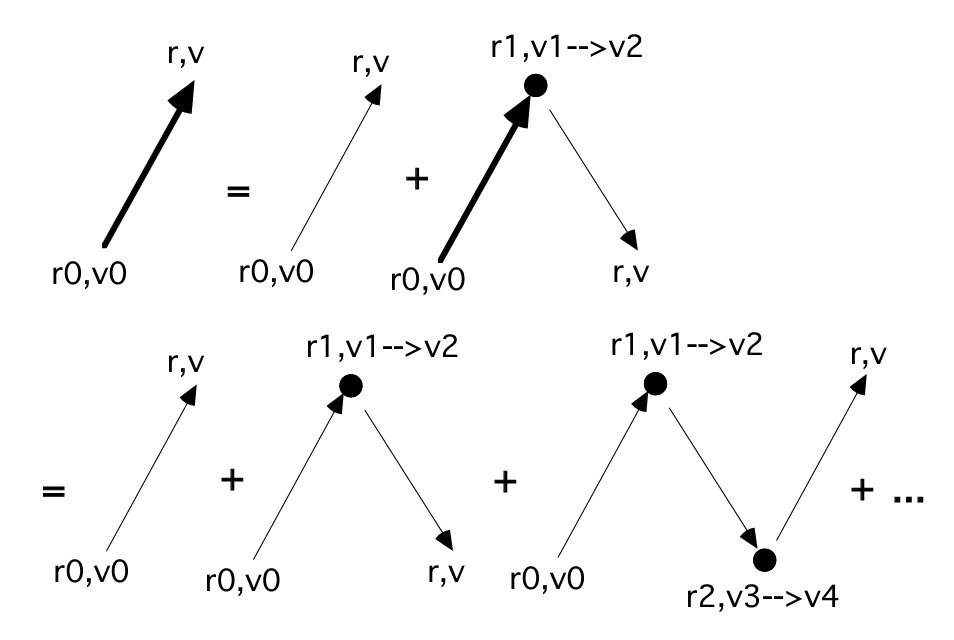}
    \caption{Representation of a MC calculation as a perturbative expansion of the exact propagator using a diagrammatic technique, explained in the text. A MC trajectory with a certain number of collisions $n_c$ is a contribution to the calculation of the diagram with $n_c$ dots. After \cite{longo2008derivation}.}
    \label{fig:diagram}
\end{figure}

The expression above is quite complex, and its subsequent orders are increasingly cumbersome, but it can be conveniently summarized by diagrams as shown in Figure \ref{fig:diagram}. 
In the figure, once again using a notation borrowed from quantum field theory \cite{mattuck1992guide}, an arrow represents a Green function, a dot represents the effect of integration and of the positive part of the collision operator $J$, a thick arrow represents the {\it exact} Green function $G$ (i.e.\ the Green function of the full transport equation). $g$, the thin arrow, represents not only collisionless particle propagation, but includes exponential decay, this is essential for the full $G$, the thick arrow, to conserve the normalization of $f$. If the null-collision method is used, the operator represented by the dot will contain a term not affecting $f$ and a term with $cJ$, $c<1$.

The factor $1/2$ which precedes the second term in Equation \eqref{eq:expansion} takes into account the fact that a collision at $t_1$ followed by a collision at $t_2$ is equivalent to the same collisions in opposite order. The term of $n$-th order will therefore be preceded by a factor $1/(n!)$. 
If the perturbation expansion is written using the null collision method, then a factor of the type $\exp(-\nu t)[\nu t]^n/(n!) $ will appear in front of each term in the expansion \cite{longo2004monte, longo2008derivation}; the details of the derivation are sketched in the next section. These factors constitute a Poisson distribution, which can be generated using the formula for collision times \eqref{eq:collision_times}. In this way, the formula normally introduced on an empirical basis finds a mathematical motivation.

These diagrams show that the null-collision MC method is in fact directly related to a specific form of the transport equation \cite{longo2002direct}. 
Indeed, it is possible to formally state that, given a set of MC simulations of a given case study, the subset that includes exactly $n$-collisions corresponds to a stochastic calculation of the term of $n$-th order in a perturbation expansion of the solution of the equation transport.

Sometimes, one can be misled by the fact that the MC method introduces statistical fluctuations of the quantities then felt in the transport equation, such as in particular the kinetic distribution of velocities. This seems to show that it has a higher element of describing reality than these equations. In reality the statistical fluctuations of the MC method are not physical and simply have to be reduced as much as possible by accumulation of events, leading to statistical convergence towards the exact solution. The method always arrives at this for an infinite number of events, when the solved equation is linear. We take advantage of this aspect to note something that is also essential in the other sections of this review, namely the fact that swarm problems are generally described by linear equations, given that the particles of the swarm are diluted to the point of not interacting. This is a basic requirement to apply the concept of the Green function.

These considerations allow us to place the MC method, normally considered a technical tool in the field of plasma simulation, in a broader context which is that of methods for theoretical physics. A very appropriate position given that it is formally capable of solving the Boltzmann transport equation, and not only as a numerical experiment, from an analogical point of view, as the terminology often used in publications seems to suggest \cite{cercignani1969mathematical, duderstadt1979transport}. 

This physical-mathematical aspect of the method can be disregarded in its most basic implementations and it can be developed precisely having in mind a direct image of the physics of the transport process.

\section{The Monte Carlo Flux method}\label{sec:monte_carlo_flux}

\subsection{Principle}
In the previous section, we noted that the traditional MC method corresponds in any case to a formal calculation, i.e.\ the calculation of the Green functions, or electron propagators. These functions are generalized functions (i.e.\ Dirac delta functions), with a diffuse component which becomes the majority as time increases, precisely due to collisions \cite{longo2008derivation}. The computed Green functions in each case depend on continuously varying arguments, which represent initial and final position and velocity components. 

Since the beginning of the '90s, modifications of the method have been experimented, in which instead of calculating propagators at the level of description of the electrons for the entire simulated time, the formalism of Markov processes has been employed to separate the diffusion time scale from the collision scale \cite{van1992stochastic}. Among these, the Monte Carlo Flux method (MCF) stands out in particular, which is a hybrid method between the deterministic method of the propagators and the stochastic MC method \cite{schaefer1990monte}. 
%These method, although very simple in the final implementation and this is their undoubted advantage, are based on more than one subtle and powerful idea and these elements must be appreciated individually in order to grasp the scope of the method itself. 
The fundamental idea of this method is the separation between diffusion and collisional time scales. In fact, due to the small mass of the electron compared with that of atoms and molecules, in some very common regimes, hundreds of collisions are needed to observe a significant average dispersion of the initial kinetic energy. It is therefore convenient to describe the diffusive phase by means of averaged quantities obtained previously from calculations describing the effect of the collisions on the single electrons. To make this calculation practical, it is advisable to carry out a coarse-grained preliminary operation by introducing a lattice of suitable geometry into the configuration space. Once this is done, diffusive kinetics over long times can be described using the Master Equation or Kolmogorov equation \cite{van1992stochastic, gardiner1985handbook} (i.e.\ the balance of probability over time). This corresponds to the Markov process in continuous time but in a discretized space. 

\begin{figure}[h]
    \centering
    \includegraphics[width=\textwidth]{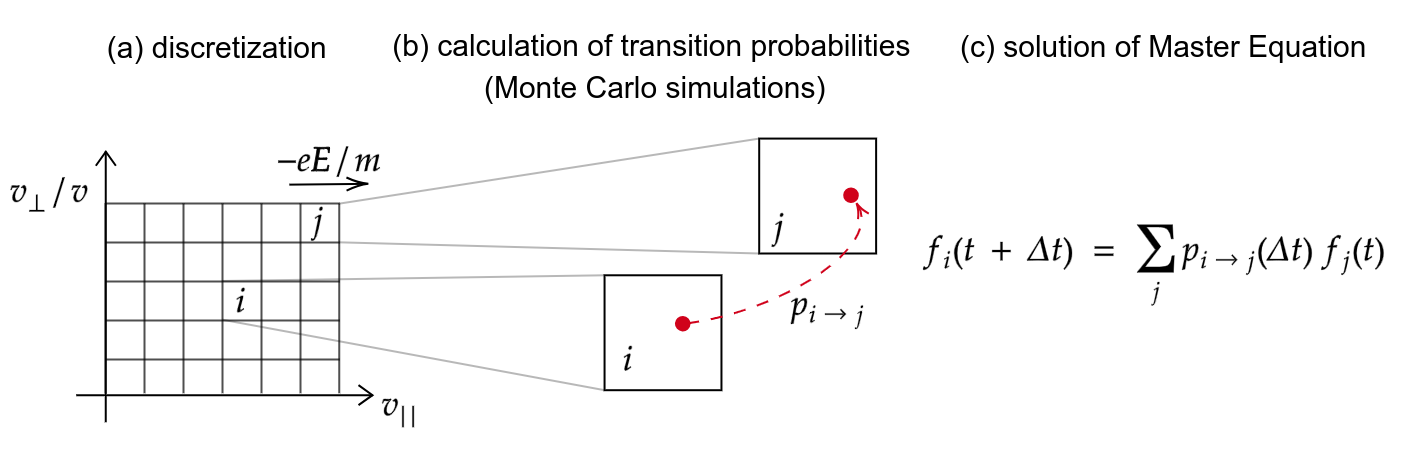}
    \caption{The MCF schematics. $(a)$ discretization of the velocity space into cells, $(b)$ MC simulations of electrons for calculation of transition probabilities between velocity space cells, $(c)$ solution of Master Equation for deterministic evolution of the electron distribution function. }
    \label{fig:mcf_scheme}
\end{figure}

The schematics of MCF is shown in Figure \ref{fig:mcf_scheme}, where the three main steps of the numerical methods are highlighted (in order from left to right).
The transition frequencies $p_{ij}(\Delta t)$ which appear as coefficients depending on two indices and on time in the equation are precisely the quantities which can be calculated by means of a collection of traditional MC simulations \cite{schaefer1990monte}. In each of these collections of simulations, electrons are uniformly distributed within a cell of the lattice in the configuration space and a simulation of the duration $\Delta t$ corresponding to a few collisions is therefore performed. At the end of the simulation, the electrons found in each of the cells estimate the transition frequency between the initial cell and the final cell. Running this collection of preliminary simulations can take significant time, as it is necessary to break down for each individual cell. However, the Markov matrix that is calculated in this way can be used as a Green function in coarse-grained space and for a fixed time difference, so that it is possible to perform different simulations, for different times as long as they are multiples of $\Delta t$, and for several initial distributions, all without the need to run the preliminary MC simulations again. Alternatively, the stationary distribution can be calculated by looking for the stationary vector with respect to the $p_{ij}$ matrix. All of these calculations are completely deterministic and pretty fast. 

\subsection{Mathematical derivation of Monte Carlo Flux}\label{subsec:mathematical}

In this subsection, the MCF method is derived 
starting from a continuous generalization and 
proceeding with its discretized form, obtained under the multigroup approximation \cite{prinja2010general}. 
This subsection is based on Chapter 3 of the PhD thesis of one of the authors \cite{vialetto2021modelling}.

The system under consideration is assumed to be homogeneous, such that spatial terms in the EBE 
can be neglected. Under these assumptions, the EBE is
\begin{equation}
\frac{\partial f(\bm{v}, t)}{\partial t} + \bm{a}(t)\cdot\bm{\nabla}_{\bm{v}}f(\bm{v}, t) = J[f],
\label{eq:boltzmann_again}
\end{equation}
\noindent
where $f(\bm{v}, t)$ is the EVDF, $\bm{a}(t)$ is the electron acceleration due to an external electric field, 
and $J[f]$ is the collision term, that in its stochastic form is 
\begin{equation}
J[f] = \int_{\bm{v}'} \left[b(\bm{v}, \bm{v}')f(\bm{v}', t) - b(\bm{v}', \bm{v})f(\bm{v}, t)\right] d\bm{v}'.
\label{eq:collision_operator}
\end{equation}
\noindent
Moreover, the following usual normalization condition applies for the transition probabilities
\begin{equation}
\nu(v) = \int_{\bm{v}'} b(\bm{v}', \bm{v}) d\bm{v}',
\label{eq:normalization_coll_freq}
\end{equation}
\noindent
with $\nu(v)$ the total electron impact collision frequency. 

A possible technique for the solution of Equation \eqref{eq:boltzmann_again} is based on the path-integral formulation, as described by Rees \cite{rees1970numerical}.
In \cite{rees1970numerical}, the rigorous path-integral theory has been initially
 developed for transport studies in semiconductors. 
 The same theory has been extended by Kumar \cite{kumar1981short} and Longo \cite{longo2002direct} 
 for the motion of charged particles in a gas in the presence of an external field.
The use of this formulation has a twofold advantage, that is; $(i)$ 
the null-collision technique \cite{skullerud1968stochastic} is directly derived from the 
transport problem and $(ii)$ the solution of Equation \eqref{eq:boltzmann_again} is obtained by
 iterative applications of a continuous generalization of a Markov operator on the initial EVDF.
In particular, from Equation \eqref{eq:collision_operator} and Equation \eqref{eq:normalization_coll_freq}, 
it is possible to verify that Equation \eqref{eq:boltzmann_again} can be rewritten as 
\begin{equation}
\frac{\partial f(\bm{v}, t)}{\partial t} + \bm{a}(t)\cdot\bm{\nabla}_{\bm{v}}f(\bm{v}, t) = M(\bm{v}, t) - \nu_{\rm max}f(\bm{v}, t),
\label{eq:boltzmann_rewitten}
\end{equation}
\noindent
where
\begin{equation}
M(\bm{v}, t) = \int_{\bm{v}'} b(\bm{v}, \bm{v}') f(\bm{v}', t) d\bm{v}' + 
\left[\nu_{\rm max} - \nu(v)\right] f(\bm{v}, t), 
\label{eq:markov_operator_M}
\end{equation}
\noindent
and $\nu_{\rm max}$ is the maximum collision frequency defined such that $\nu_{\rm max} \geq \nu(v)$ for all values of the 
electron speed $v$. 
Note that the first term on the right-hand-side of Equation \eqref{eq:markov_operator_M} 
represents the gain of electrons having velocity $\bm{v}$ at time $t$ due to collisions with the 
background gas, 
and the term $(\nu_{\rm max} - \nu(v))$ represents ``null-collision'' events per 
unit time that leave the EVDF unchanged.  
The definition of $\nu_{\rm max}$ is also widely used in MC simulations, since it is at the basis 
of the null-collision technique introduced by Skullerud \cite{skullerud1968stochastic}.
For stationary conditions the solution of Equation \eqref{eq:boltzmann_rewitten} is
\begin{equation}
f(\bm{v}) = \int_0^{\infty} M(\bm{v} - \bm{a}t) e^{-\nu_{\rm max}t}dt.
\label{eq:stationary_solution}
\end{equation}
\noindent
The method for evaluating the right-hand-side of Equation \eqref{eq:stationary_solution} is based on the following 
iterative technique \cite{rees1970numerical}. 
First, an intermediate function $M_{(n-1)}(\bm{v})$ for the $(n-1)$-th iteration is generated from $f_{(n-1)}(\bm{v})$, as 
\begin{equation}
M_{(n-1)}(\bm{v}) = \int_{\bm{v}'} b(\bm{v}, \bm{v}') f_{(n-1)}(\bm{v}') d\bm{v}' + [\nu_{\rm max} - \nu(v)]f_{(n-1)}(\bm{v})
\label{eq:iterative_M}
\end{equation}
\noindent
Second, the distribution $f_{(n)}(\bm{v})$ is obtained from $M_{(n-1)}(\bm{v})$ as
\begin{equation}
f_{(n)}(\bm{v}) = \int_0^{\infty} M_{(n-1)}(\bm{v} - \bm{a}t) e^{-\nu_{\rm max}t} dt, 
\label{eq:iterative}
\end{equation}
\noindent
The advantage of this method is that the $n$-th iterate $f_{(n)}(\bm{v})$ is equivalent to the 
distribution obtained after a time interval $n/\nu_{\rm max}$.
Furthermore, the stationary EVDF can be obtained in the limit 
$f(\bm{v}) = \lim_{n\to\infty} f_{(n)}(\bm{v})$.

As described by Nanbu \cite{nanbu1980direct}, the same iterative approach can be applied when 
the temporal evolution of the EVDF is sought after.
In particular, for sufficiently high values of $\nu_{\rm max}$, let $\Delta t = (\nu_{\rm max})^{-1}$ be the time 
step for EVDF evolution. Then, at first order in $\Delta t$, the EVDF can be expressed as
\begin{equation}
f_{(n)}(\bm{v}) \simeq M_{(n-1)}(\bm{v} - \bm{a}\Delta t)\Delta t,
\label{eq:time_evol}
\end{equation}
\noindent
or equivalently
\begin{equation}
f_{(n)}(\bm{v}) \simeq \Delta t \int_{\bm{v}'} b(\bm{v}, \bm{v}') f_{(n-1)}(\bm{v}') d\bm{v}' 
+ [1 - \nu(v)\Delta t] f_{(n-1)}(\bm{v} - \bm{a}\Delta t).
\label{eq:iterative_time_evolution}
\end{equation}
\noindent
If $\Delta t$ is sufficiently low, an iterative solution of Equation \eqref{eq:boltzmann_rewitten} describes also 
the temporal evolution of the EVDF, starting from an initial distribution $f_{(0)}$, at time $t=0$. 

To summarize, it is useful to represent the aforementioned equations in a schematic form. 
In particular, the integration on the right-hand-side of Equation \eqref{eq:iterative_M} can be represented as 
the action of an operator $\hat{S}$ on $f_{(n-1)}(\bm{v})$ and that on the right-hand-side of Equation \eqref{eq:iterative} by 
the action of an operator $\hat{P}$ on $M_{(n-1)}(\bm{v})$. 
A complete iteration is therefore described by the action 
of the combined operator $\hat{M} = \hat{S}\hat{P}$ on $f_{(n-1)}(\bm{v})$.
Moreover, the time evolution of an initial function $f_{(0)}(\bm{v})$ after a time interval $n/\nu_{\rm max}$ 
is determined by $\{\hat{M}\}^n$.
Different numerical techniques can be employed for calculations of the aforementioned operator. 
In the MCF method, a grid is defined for calculations of transition probabilities for the electron motion between velocity space cells. 
In this discretized form, $\hat{M}$ is also termed as the transport matrix (or Markov matrix). 
The following subsections are devoted to the description of this discretization and the numerical procedure that is applied for calculations of $\hat{M}$.

\subsection{Discretization of the Electron Velocity Distribution Function}

We describe here the numerical method that is used to implement the MCF. 
A more detailed comparison of MCF against other methods, such as conventional MC, 
two-term Boltzmann solver, and multi-term Boltzmann solver, is described in 
\cite{schaefer1990monte, vialetto2019benchmark, vialetto2020benchmarking}. Furthermore, 
the method has been recently coupled with detailed plasma chemistry models in 
\cite{viegas2020insight, viegas2021resolving, micca2021plasma, vialetto2022charged}.

In this subsection, a uniform external electric field $\bm{E}$ is considered, such that $\bm{E} = (0, 0, - | E_z |)$. 
Hence the electron velocity $\bm{v} = (v_x, v_y, v_z)$ can be represented in polar coordinates $(\epsilon, \theta, \phi)$, 
where $\epsilon$ is the electron kinetic energy (i.e.\ $\epsilon = (1/2) m v^2$), $\theta$ is the angle between $\bm{v}$ and the 
$z$-direction, and $\phi$ is the azimuthal angle around the $z$-axis.
Under the assumption of isotropic scattering around the $z$-axis, the velocity space can be described by 
only two variables $(\epsilon, \theta)$, instead of three. Moreover, under this assumption, the EVDF is uniform in $\phi$ (i.e.\ $f(\bm{v}) = f(\epsilon, \cos\theta)$). 
Hence the velocity space $(\epsilon, \cos\theta)$ is partitioned into cells $C_{i, j}$, 
with $1 \leq i \leq I$ the index for the energy component and $1 \leq j \leq J$ the index for the angular component.
The calculation range is assumed as $0 \leq \epsilon \leq \epsilon_{\rm max}$ and $-1 \leq \cos\theta \leq 1$, where 
$\epsilon_{\rm max}$ is chosen such that we can neglect electron diffusive fluxes to energies higher than $\epsilon_{\rm max}$. In this way, the energy and angular bin sizes are defined as $\Delta\epsilon = \epsilon_{\rm max}/I$ and 
$| \Delta(\cos\theta) | = 2/J$.  

Under these assumptions, 
the EVDF can be represented in its discretized form as a column vector $\bm{f}$ of size $IJ \times 1$ as
\begin{align}\label{eq:evdf_col_vect}
    \bm{f} &= \begin{bmatrix}
           \bm{n}_{1} \\
           \bm{n}_{2} \\
           \vdots \\
           \bm{n}_{J}
         \end{bmatrix},
  \end{align}

\noindent
where each element of the vector $\bm{f}$ has size $I \times 1$ and represents the 
total number of particles having energies between 0 and $\epsilon_{\rm max}$ and $\cos\theta = \cos\theta_j$, 
with $j = 1, \dots, J$. 
In particular, the $j$-th component ($\bm{n}_j$) is written as

\begin{align}\label{eq:n_col_vect}
    \bm{n}_j &= \begin{bmatrix}
           n_{1, j} \\
           n_{2, j} \\
           \vdots \\
           n_{I, j}
         \end{bmatrix},
  \end{align}
  
\noindent
where $n_{i, j}$ is the total number of electrons in the $C_{i,j}$ cell and it is computed as

\begin{equation}
n_{i, j} = \sum_{k=1}^{N_{\rm elec}} \delta(\epsilon_k) \delta(\cos\theta_k), 
\label{eq:components_n}
\end{equation}

\noindent
where $N_{\rm elec}$ is the total number of particles in the MC simulation and 

\begin{equation}
    \delta(\epsilon_k) = 
\begin{cases}
    1,& \text{if } (i - 1)\Delta\epsilon \leq \epsilon_k \leq i \Delta\epsilon\\
    0,              & \text{elsewhere}
\end{cases},
\label{eq:delta_eps}
\end{equation}

\noindent
\begin{equation}
    \delta(\cos\theta_k) = 
\begin{cases}
    1,& \text{if } -1 + j\Delta |\cos\theta| \leq \cos\theta_k \leq -1 + (j - 1) \Delta |\cos\theta| \\
    0,              & \text{elsewhere}
\end{cases}.
\label{eq:delta_cos}
\end{equation}

\noindent
In other words, after discretization of the energy and angular components, 
each simulated particle contributes to the EVDF as a Kronecker delta function \cite{yousfi1994monte}.
From the EVDF, the $l$-th order Legendre polynomial coefficients $f_l(\epsilon_i)$
can be calculated as

\begin{equation}
f_l(\epsilon_i) = A_l(\epsilon_i) \bm{f}^{\top}(\epsilon_i) \bm{L}_l,
\label{eq:leg_pol}
\end{equation}

\noindent
where $A_l(\epsilon_i) = (2l + 1)/(n_{\rm e} \Delta\epsilon \sqrt{\epsilon_i})$ is a normalization factor that depends 
on the total electron population $n_{\rm e}$, such that

\begin{equation}
n_{\rm e} = \sum_{i=1}^I \sum_{j=1}^J n_{i, j},
\label{eq:electron_density_normalization}
\end{equation}

\noindent
$\bm{f}^{\top}(\epsilon_i)$ is the transpose vector for the EVDF computed at $\epsilon = \epsilon_i$, 
thus having size $1 \times J$, and $\bm{L}_l$ is a column vector of size $J \times 1$ written as

\begin{align}
    \bm{L}_l &= \begin{bmatrix}
           P_l(\cos\theta_1) \\
           P_l(\cos\theta_2) \\
           \vdots \\
           P_l(\cos\theta_J)
         \end{bmatrix},
\end{align}

\noindent
with $P_l(\cos\theta_j)$ the $l$-th order Legendre polynomial calculated at $\cos\theta_j$.
Note that the Electron Energy Distribution Function (EEDF) 
comes from Equation \eqref{eq:leg_pol} as the zero-th order (isotropic) component ($f_0$) of the EVDF expansion. 
Hence $f_0$ can be calculated with the following simplified expression:

\begin{equation}
f_0(\epsilon_i) = \frac{1}{n_{\rm e} \Delta\epsilon \sqrt{\epsilon_i}} \sum_{j=1}^J n_{i,j},
\label{eq:eedf_simple}
\end{equation}

\noindent
meaning that the EEDF can be retrieved directly from binning in energy of the total number of simulated particles. 
In the following, the basic idea underlying MCF simulations is covered, that is the calculation of the temporal evolution 
of the EVDF, given an initial distribution at time $t = 0$.

\subsection{Calculation of the transport matrix}
\label{ch3:subsec:ConstructionTransitionProb}

In the MCF method, the temporal evolution of the EVDF is obtained by 
a matrix-vector operation, given an initial distribution function at the time $t=0$. 
The EVDF at time $t + \Delta t$ is computed as \cite{schaefer1990monte}

\begin{equation}
\bm{f}(t + \Delta t) = \bm{M}(\Delta t)^{\top} \bm{f}(t),
\label{eq:time_evolution_evdf}
\end{equation}

\noindent
where $\bm{M}(\Delta t)$ is the transport matrix of size $IJ \times IJ$.  
Note that the matrix $\bm{M}$ is a discrete version of the 
continuous operator $\hat{M}$ defined in Section \ref{subsec:mathematical}. 
%Hence, elements of $\bm{M}$ 
%are determined by MC simulations that trace the electron motion between 
%velocity space cells due to collisions with the background gas and acceleration from 
%the external electric field.

More generally, 
if the reduced electric field and the gas composition are fixed for any $t > 0$, 
the transport matrix remains unchanged  
and the EVDF can be found with an iterative procedure as

\begin{equation}
\bm{f}(t + n\Delta t) = \left(\bm{M}(\Delta t)^{\top}\right)^n \bm{f}(t),
\label{eq:time_evolution_iterative}
\end{equation}

\noindent
with $n$ the iteration number, and $n\Delta t$ the time-step associated with the $n$-th iteration. 
An important consequence of equation \eqref{eq:time_evolution_iterative} is that the evolution of the system after $\Delta t$ is determined only by the state of the system at a time $t$ and it is not affected by the previous history. This is known as Markov property and allows one to rewrite the linear EBE (Equation \eqref{eq:boltzmann_again}) as a simple Markov chain consisting of the system of linear equations in \eqref{eq:time_evolution_iterative}. This property is typically satisfied if $t_{coll} \ll \Delta t < t_{SS}$, that is $\Delta t$ should be much longer than the time interval between two successive collisions $t_{coll}$, but also shorter than the time $t_{SS}$ for the distribution function to reach steady-state.  
In particular, collisions are essential for the randomization of the particle velocities and trajectories. Through this randomization, electron history is erased and the evolution of the system depends only on the current state, not on past states. Hence, as shown in \cite{vialetto2019benchmark, vialetto2020benchmarking}, the choice of an appropriate $\Delta t$ 
is fundamental to ensure the applicability of Equation \eqref{eq:time_evolution_iterative}. 

The form of the transport matrix is 
\begin{equation}
\bm{M} = 
\begin{bmatrix}
\bm{m}_{1,1} & \bm{m}_{1,2} & \cdots & \bm{m}_{1,J} \\
\bm{m}_{2,1} & \bm{m}_{2,2} & \cdots & \bm{m}_{2,J} \\
\vdots  & \vdots  & \ddots & \vdots  \\
\bm{m}_{J,1} & \bm{m}_{J,2} & \cdots & \bm{m}_{J,J} 
\end{bmatrix}
\label{eq:matrix_bigM}
\end{equation}
\noindent
where $\bm{m}_{j,j'}$ are sub-matrices of size $I \times I$. 
The total number of sub-matrices in Equation \eqref{eq:matrix_bigM}
is determined by the angular discretization. For example, if $J$ cells are defined for the 
angular component, then all the $J \times J$ possible combinations are included in $\bm{M}$.
Each sub-matrix is written as:
\begin{equation}
\bm{m}_{j,j'} = 
\begin{bmatrix}
p_{1, 1} & p_{1,2} & \cdots & p_{1,I} \\
p_{2,1} & p_{2,2} & \cdots & p_{2,I} \\
\vdots  & \vdots  & \ddots & \vdots  \\
p_{I,1} & p_{I,2} & \cdots & p_{I,I} 
\end{bmatrix}_{j, j'},
\label{eq:matrix_smallM}
\end{equation}
\noindent
where $p_{i, i' | j, j'}(\Delta t)$ is the transition probability 
for electrons moving from cell $C_{i, j}$ to $C_{i', j'}$ 
within the time interval $\Delta t$. 
In the MCF method, short MC simulations are performed for calculations of transition probabilities 
between all velocity space cells. Hence, 
transition probabilities in Equation \eqref{eq:matrix_smallM} are calculated as 
\begin{equation}
p_{i, i' | j, j'}(\Delta t) = n_{i', j'}(\Delta t)/n_{i, j}(0),
\label{eq:transition_prob}
\end{equation}
\noindent
where $n_{i', j'}(\Delta t)$ is the total number of electrons moving from cell $C_{i, j}$ to $C_{i', j'}$ within the time 
interval $\Delta t$ and $n_{i, j}(0)$ is the 
total number of electrons in the initial cell $C_{i, j}$ at time $t=0$.
Note that the time interval $\Delta t$ depends on physical parameters 
such as the electric field $E$ and the 
gas number density $N$. 
Empirically, it becomes shorter at higher $E$ and $N$, as the electron collision frequency increases. 
The choice of an optimal $\Delta t$ is important for an optimization of the CPU time 
in MC simulations, as well as for accurate calculations of transition probabilities \cite{vialetto2019benchmark}.

\subsection{Advantages and disadvantages}

The main advantages of the MCF method are the following \cite{vialetto2019benchmark, vialetto2020benchmarking, vialetto2021modelling}:
\begin{itemize}
    \item Since the MCF is used to calculate the transition frequency and not directly the kinetic distribution, the MCF results have uniform statistical fluctuations for all the regions of the distribution, in particular also the tail, which instead may be statistically inaccessible to the traditional method given the low number of electrons described.
\item The matrix-based approach does not require a series expansion of the EVDF and
it is easier to implement than a multi-term Boltzmann solver. In perspective, the
MCF method can be combined with efficient algorithms for matrix operations
and GPU acceleration.
\item As a difference with respect to other variance reduction techniques, mainly
based on variable mathematical weights for the simulated particles, the MCF
method addresses the fundamental problem of the very large ratio, amounting
to several orders of magnitude, between the relaxation time of the distribution
and the inter-collision time. Hence, the stochastic part of MC simulations is
limited to a small time interval, that is typically orders of magnitude lower than
the steady-state time for the electron energy distribution function (EEDF).
\end{itemize}

However, in spite of the advantages of this method in terms of both computational cost
and accuracy with respect to a conventional MC method, a
number of disadvantages of MCF should be considered when choosing
an appropriate computational method for plasma applications. For example, 

\begin{itemize}
    \item The increasing size of the transport matrix lengthens the computational time
and this is also a limit for practical use of the MCF method. This is a problem,
for example, for an extension to the configuration space, since additional calculations of transition probabilities of electrons moving between different cells
in the spatial coordinates are needed. Nevertheless, matrices in this case are
largely sparse. Hence, the use of efficient algorithms for sparse matrix calculations could help to reduce the large memory that is required.
\item The method presented here can only deal with calculations
of flux transport parameters. However, bulk parameters are needed when comparing results of calculated electron transport coefficient with swarm measurements, especially at high E/N \cite{petrovic2009measurement}. An extension of the MCF method to the
configuration space will provide calculations of the aforementioned parameters
as well.
\item The transport matrix includes the effect of both field and collisional events. This has practical limitations for calculations of transition probabilities in the
presence of a time-varying electric field evolving in timescales comparable with
the energy relaxation time. 

\end{itemize}

Some of the limitations of the MCF method can be overcome by the Propagator method that is described in the following Section, where the the field acceleration is separated from the collision part by defining different propagator matrices similar to the transport matrices used in MCF. This makes the Propagator method also applicable to AC electric fields. However, the  the Propagator method is limited to low $\Delta t$ such that the electron out flow from a cell does not exceed the total number of electrons in that cell \cite{Sugawara2017}. Furthermore, it is important to highlight that, as a difference with respect to the Propagator method and the Convection scheme \cite{hitchon1989efficient}, the MCF uses MC simulations for calculations of transition probabilities between velocity space cells. The use of MC simulations is advantageous for its simpler implementation with respect to efficient convective schemes, but it is computationally expensive for large domain simulations. 

\section{Propagator method}\label{sec:propagator_method} % //// Sugawara part ////

\subsection{Principle}

Propagator method (PM) is a numerical scheme to calculate
 the EVDF (or EEDF) on the basis of the EBE\@.
The PM shares many common concepts with the MCF method.
The PM calculation is performed with cells
 defined by partitioning of phase space $(\bm{v}, \bm{r})$,
 velocity space and real space,
 into small sections in which electrons distribute.
The number of electrons belonging to a cell is stored
 in an element of a matrix or an array corresponding to the cell.
The probability of intercellular electron transition
 from a source cell to destination cells in a short time step $\Delta t$
 is represented by operators called propagator or the Green function.
The transition may be caused by
 the velocity change under electric and magnetic fields
 and scattering at collisions with gas molecules.
The PM does not use random numbers,
 and its calculation is completely deterministic.
The stochastic processes are condsidered
 on the bases of their expected values.
Temporal development of the EVDF is observed
 by applying the acceleration and collision propagators
 to the EVDF every $\Delta t$.
The PM has a flexibility in treatment of the propagators
 as seen in practical examples introduced in the following sections.
The step-by-step time development of the EVDF can be modified
 in some specific models
 customized to derive only equilibrium solutions,
 and the PM can deal with some simple boundary conditions
 such as electron reflection and secondary electron emission
 at electrode surfaces.

There are various cell configurations and experimental models
 for the PM calculations
 depending on the target properties of electron swarms.
Let us overview efforts of calculations
 which have ever been performed using the PM\@.

\subsection{Classification of cell configurations and expressions of electron motion}

We may classify the types of the PM configurations
 on the basis of the partitioning of velocity space for cells and
 the expressions of the electron motion
 in quantification of the intercellular electron transition.

\begin{figure}[t] \begin{center} % -------------------------------------
\includegraphics[keepaspectratio, width=120mm, clip]{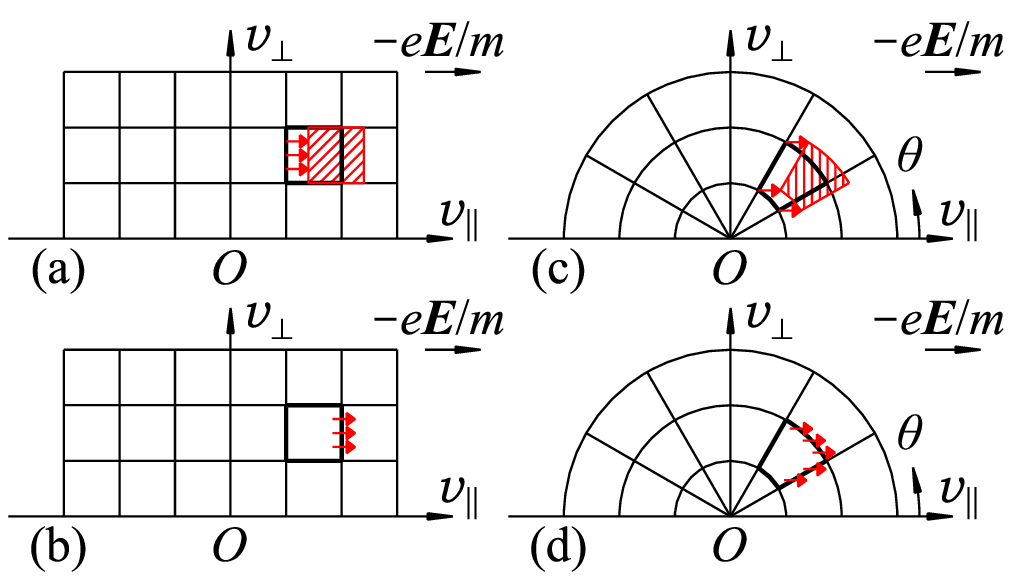}
\end{center} \caption{Cells defined in two-variable velocity space
 $(v_\parallel, v_\perp)$ or $(v, \theta)$,
 and treatment of the intercellular electron transition by acceleration:
 a, Lagrangian treatment in Cartesian-$v$ cell configuration;
 b, Eulerian treatment in Cartesian-$v$ cell configuration;
 c, Lagrangian treatment in polar-$v$ cell configuration; and
 d, Eulerian treatment in polar-$v$ cell configuration.
The thick boundaries indicate the source cells
 from which electrons flow out downstream (rightward) by acceleration.
The red hatched cells are the Lagrangian cells.
} \label{fig:cell-acc} \end{figure}

For the former point, the cells can be defined, for example,
 for every $\Delta v_x$, $\Delta v_y$, and $\Delta v_z$
 in Cartesian coordinate system $(v_x, v_y, v_z)$
 (Figures \ref{fig:cell-acc}a and \ref{fig:cell-acc}b),
 or for every $\Delta v$, $\Delta \theta$, and $\Delta \phi$
 in polar coordinate system $(v, \theta, \phi)$
 (Figures \ref{fig:cell-acc}c and \ref{fig:cell-acc}d),
 where $v_z = v \cos \theta$, $v_x = v \sin \theta \cos \phi$,
 and $v_y = v \sin \theta \sin \phi$.
As well, division for every $\Delta \epsilon$ instead of $\Delta v$
 is also possible.
They are referred to as
 Cartesian-$v$, polar-$v$, and polar-$\epsilon$ configurations of
 the cells for velocity space in the following sections.

For the latter point, Lagrangian and Eulerian expressions are considered
 for the treatment of electron motion in the acceleration or flight.
In the Lagrangian expression,
 the source and destination cells are related
 by ballistic motion of electrons starting from the source cell
 (Figures \ref{fig:cell-acc}a and \ref{fig:cell-acc}c),
Think that the electrons in a source cell undergo
 a collective free flight for $\Delta t$.
They would appear as if they are a moving cell,
 which is called a Lagrangian cell.
The destination cells are chosen as those located at the position
 where the Lagrangian cell reaches after $\Delta t$,
 and the ratio of the electron redistribution
 that the destination cell accepts from the source cell
 is calculated as the ratio of the area
 overlapping between the Lagrangian cell and each destination cell.
On the other hand, in the Eulerian expression,
 source and destination cells are assumed to contact each other
 and to share the cell boundary between them.
The number of electrons
 moving out of the source cell to the destination cell,
 $n_{\rm e,out}$, is evaluated by integrating the electron flux
 passing through the cell boundary during $\Delta t$
 (Figs. \ref{fig:cell-acc}b and \ref{fig:cell-acc}d).
Its amount is evaluated as $n_{\rm e,out}
 = S_{\rm cell} (n_{\rm e,cell} / V_{\rm cell}) (e E/m) \Delta t$,
 where $S_{\rm cell}$ [${\sf L}^2 {\sf T}^{-2}$]
 is the area of the cell boundary
 projected to a plane perpendicular to $\bm{E}$,
 $n_{\rm e,cell}$ is the number of electron in the cell,
 $V_{\rm cell}$ [${\sf L}^3 {\sf T}^{-3}$] is the volume of the cell,
 and $(e E/m)$ [${\sf L} {\sf T}^{-2}$] is the electron acceleration.
$\Delta t$ must be short enough to avoid
 $n_{\rm e,out} > n_{\rm e,cell}$ for all cells.

 % ----------------------------------

\subsection{Collision propagator}

\begin{figure}[t] \begin{center} % -------------------------------------
\includegraphics[keepaspectratio, width=120mm, clip]{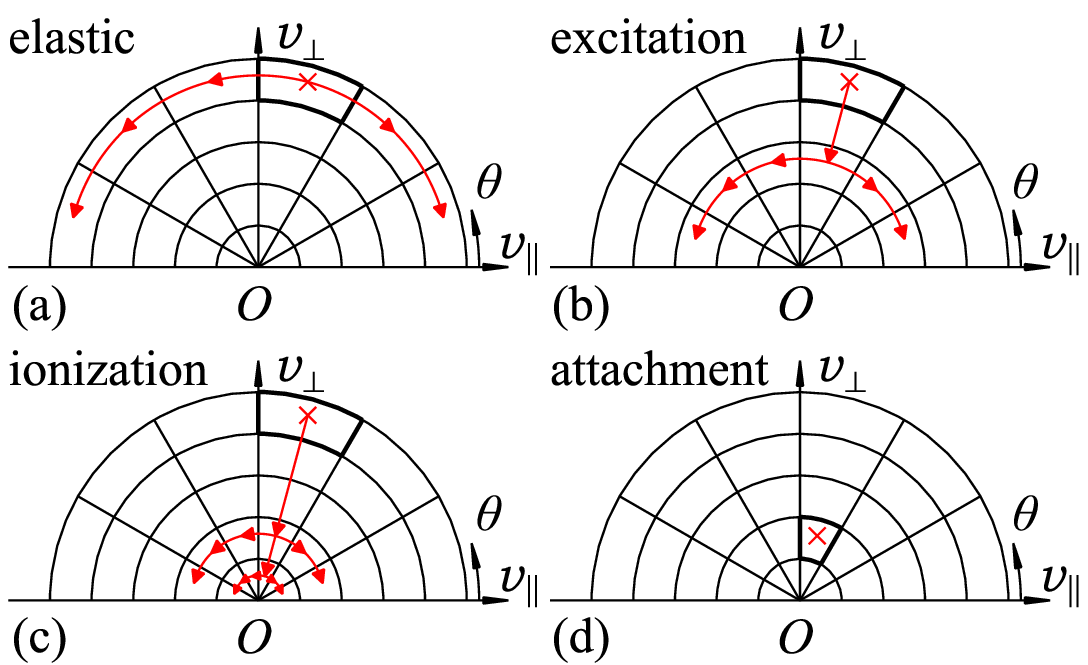}
\end{center} \caption{Treatment of electron collisions and scattering:
 a, elastic collision,
 electron redistribution to the cells of $\epsilon = \epsilon'$;
 b, excitation collision,
 electron redistribution to the lower-energy cells of
 $\epsilon = \epsilon' - \epsilon_{\rm exc}$;
 c, ionization collision,
 electron doubling and redistribution to the lower-energy cells of
 $\epsilon \le \epsilon' - \epsilon_{\rm ion}$; and
 d, electron attachment, electron disappearance from velocity space.
The cells with thick boundaries and red crosses are the source cells,
 from which electrons flow out.
$\epsilon'$ is the energy of the incident electron.
$\epsilon_{\rm exc}$ and $\epsilon_{\rm ion}$ are
 the excitation and ionization energies, respectively.
} \label{fig:cell-coll} \end{figure}

The collision propagator represents the change of electron velocity
 due to collision and scattering.
The collisions are categorized into mainly four types.
In the simplest case, the following processes are assumed
 for the electrons undergoing collision at an energy $\epsilon'$.
\begin{description}
\item[Elastic collision]
 The electrons are scattered isotropically without loss of energy.
 They are redistributed to the destination cells
 having the same energy $\epsilon = \epsilon'$ in proportion to
 the solid angle of the destination cell
 subtended at the origin of velocity space
 (Figure \ref{fig:cell-coll}a).
\item[Excitation collision]
 The electrons lose excitation energy $\epsilon_{\rm exc}$
 and are redistributed to the lower-energy cells of
 $\epsilon = \epsilon' - \epsilon_{\rm exc}$
 (Figure \ref{fig:cell-coll}b).
\item[Ionization collision]
 The electrons lose ionization energy $\epsilon_{\rm ion}$ and
 the residual energy is shared by the primary and secondary electrons.
 The electrons, which are doubled,
 are redistributed to the lower-energy cells of
 $\epsilon \le \epsilon' - \epsilon_{\rm ion}$
 with relevantly given ratios under the law of energy conservation
 (Figure \ref{fig:cell-coll}c).
\item[Electron attachment]
 The electrons captured by gas molecules disappear from velocity space
 (Figure \ref{fig:cell-coll}d).
\end{description}

The number of electrons
 undergoing collisions of the $k$th kind in a cell during $\Delta t$,
 $n_{{\rm e,coll},k}$, is evaluated as
 $n_{{\rm e,coll},k} = [N q_k(v) v \Delta t] n_{\rm e,cell}$,
 where $N$ is the gas molecule number density, and
 $q_k(v)$ is the electron collision cross section of
 the $k$th kind of collisional process
 as a function of the electron speed $v$.
The portions $n_{{\rm e,coll},k}$
 are subtracted from $n_{\rm e,cell}$ of the source cell,
 and distributed to the destination cells
 corresponding to the electron velocity after scattering.
$\Delta t$ must be short enough so that the probability of
 multiple collisions during $\Delta t$ can be neglected,
 and to satisfy $\sum_k n_{{\rm e,coll},k} < n_{\rm e,cell}$.

 % ---------------------------------

\subsection{Models of velocity-space under uniform electric fields}

The PM calculation
 for the temporal development of the EVDF of an electron swarm
 under a uniform electric field $\bm{E} = (0, 0, E_z)$
 is the simplest self-consistent model.
It assumes an electron swarm in boundary-free real space
 and the position of each electron is ignored.
The $\bm{E}$ field may be dc, ac (typically rf: radio frequency),
 or even impulse fields.

The EBE for this model is written as
 \begin{align}
 \frac{\partial}{\partial t} f(\bm{v}, t)
 &=
  -  \bm{a} \cdot \frac{\partial}{\partial \bm{v}} f(\bm{v}, t)
  +  J[ f(\bm{v}, t) ] \\
 &=
  -  a_\parallel \frac{\partial}{\partial v_\parallel} f(\bm{v}, t)
  +  J[ f(\bm{v}, t) ].
 \end{align}
The spatial electron distribution has already been integrated
 and its profile is not cared here.

The EVDF in this model can be assumed to be in a rotational symmetry
 for the azimuthal direction $\phi$ around the $v_z$-axis.
Thus, the EVDF can be represented as a two-variable distribution
 as $f(v_\parallel, v_\perp, t)$ or $f(v, \theta, t)$.
Here, $v_\parallel = v_z = v \cos \theta$
 and $v_\perp = \sqrt{v_x^2 + v_y^2} = v \sin \theta$
 are components of $\bm{v}$
 parallel and perpendicular to the direction of $-\bm{E}$, respectively,
 and $v = |\bm{v}|$.

Some of the earliest PM efforts to obtain $f(\bm{v}, t)$
 adopted the Cartesian-$v$ configuration
 with $\Delta v_\parallel$ and $\Delta v_\perp$
 under dc \cite{Drallos1988, Drallos1989} and
 rf \cite{Maeda1994} $\bm{E}$ fields.
Such a cell configuration simplifies
 the calculation of the electron acceleration
 because it is a parallel shift to the $+v_\parallel$ direction
 in velocity space.
Each cell has a boundary in contact with its downstream neighbour,
 and the boundary is normal to $\bm{a}$.
The probability of electron transition from a cell
 can be calculated easily in both Eulerian and Lagrangian manners.
The Cartesian-$v$ configuration has also been adopted
 in calculation of $f(\bm{v}, z, t)$
 extended to a parallel-plane electrodes model
 including one-dimensional real space $z$
 \cite{Sommerer1989, Sommerer1991, Parker1993}.
In the one-dimensional real-space model,
 there are different treatments
 for simultaneous changes of $\bm{v}$ and $z$ in an electron flight,
 which is explained in the next subsection.

On the other hand, polar-$v$ or polar-$\epsilon$ configuration
 with $\Delta v$ or $\Delta \epsilon$,
 and $\Delta \theta$ or $\Delta(\cos \theta)$
 makes the redistribution of electrons after collisions simple
 when isotropic electron scattering is assumed.
The ratio of the scattered electrons that a destination cell receives
 is proportional to the solid angle of the cell
 subtended at the origin $\bm{v} = 0$ of velocity space
 in case of isotropic scattering,
 and the sold angle $\Omega$ of a cell is given from
 ${\rm d}\Omega = 2 \pi {\rm d}(\cos \theta)
 = 2 \pi \sin \theta {\rm d}\theta$.
Such a polar-$\epsilon$ configuration has been adopted
 not only for $f(\bm{v}, t)$
 in velocity space in pulsed Townsend (PT) mode
 \cite{Sugawara2006, Sugawara2017}
 but also for the steady-state Townsend (SST) mode
 \cite{Sugawara1992, Sugawara1994, Sugawara1995},
 for the time-of-flight (TOF) mode
 \cite{Sugawara1997, Sugawara1998, Sugawara2017},
 and rf \cite{Kobayashi2023} and impulse \cite{Sugawara2003} fields
 as mentioned afterward.

\subsection{Models of real space between parallel-plane electrodes}

Phase space to consider the EVDF
 in one-dimensional real space between parallel-plane electrodes
 is required to be three-variable
 for example as $(v_\parallel, v_\perp, z)$ or $(v, \theta, z)$.
In the point of required memory capacity for practical calculations,
 it is beneficial that
 the range of $z$ is finite between the electrodes.
The computational array for the cells becomes three-dimensional,
 and cells are prepared for every
 $\Delta v_\parallel$, $\Delta v_\perp$, and $\Delta z$
 \cite{Sommerer1989, Sommerer1991, Parker1993}, or
 for every $\Delta \epsilon$, $\theta$, and $\Delta z$
 \cite{Sugawara1992}.
The same collision propagator as used in the velocity-space model
 can be applied to this mode
 because the scattering changes only $\bm{v}$ of electrons
 and their positions are unchanged at that moment.
The acceleration propagator becomes to involve the spatial displacement
 because $\bm{v}$ and $z$ change at the same time
 in electron free flight under an electric field.

In some early PM efforts using cells defined for every
 $\Delta v_\parallel$, $\Delta v_\perp$, and $\Delta z$
 \cite{Sommerer1989, Sommerer1991, Parker1993},
 the destination cells were chosen under a concept of Lagrangian cell.
The choice of destination cells and
 evaluation of the electron redistribution ratios for them
 are easy in the Cartesian-$v$ configuration
 because the geometrical shape of cells is simple.
This treatment also arrows flexible change of $\bm{E}$
 not only for dc $\bm{E}$
 but also for rf $\bm{E}$ \cite{Sommerer1991, Parker1993}
 or position-dependent $\bm{E}$ \cite{Sommerer1989, Sommerer1991}.

On the other hand, treatment of electron displacement
 in an Eulerian manner was also attempted \cite{Sugawara1992}
 adopting restrictions
 for the cell configuration and the intercellular electron motion.
The cells are defined
 for every $\Delta \epsilon$, $\Delta \theta$, and $\Delta z$
 to satisfy $\Delta \epsilon = e E \Delta z$.
The number of electrons flowing out of a source cell is evaluated
 by integrating the electron flux
 passing through the downstream cell boundary in velocity space.
For the electrons which move from the source cell
 defined at an $\epsilon$ region
 to destination cells at the $\epsilon \pm \Delta \epsilon$ regions,
 the spatial displacement $\pm \Delta z$ accompanies, respectively.
The changes of $\epsilon$ and $z$ are strictly bound
 under the law of conservation of energy.
A demonstration was made under a SST condition
 by superposing a temporal development of
 an isolated electron swarm starting from the cathode
 with an initial energy $\epsilon = 0$
 until almost all electrons are absorbed by the anode.
Under the restrictions to guarantee the energy conservation,
 position-dependent profiles of spatial relaxation processes of
 mean electron energy and drift velocity
 agreed with results of a MC simulation.
Especially, the rising position of ionization coefficient
 was successfully reproduced at $z = \epsilon_{\rm ion}/(e E)$,
 where $\epsilon_{\rm ion}$ is the ionization threshold,
 by suppressing the numerical diffusion which may occur
 among the destination cells in the Lagrangian manner.

The boundary condition at the electrodes is also a point of discussion.
The simplest model assumes perfect absorption,
 where the electrons reaching the electrodes disappear.
Elastic or inelastic reflection at the electrodes can be considered
 by relevantly choosing $\bm{v}$
 after reaching either of the electrodes.
The effect of secondary electron emission \cite{Sommerer1991}
 can be considered as a practical condition as well
 by supplying new electrons to the cells near the electrode
 on which primary electrons impinge.

\subsection{Models of boundary-free real space in steady-state Townsend condition}

When SST condition is assumed
 for one-dimensional electron flow in $z$ direction
 under uniform electric field $\bm{E} = (0, 0, -E)$,
 partitioning of the $z$ position can be omitted \cite{Sugawara1994}
 in the PM calculation for the EVDF by utilizing the assumptions
 that the normalized EVDF is identical irrespective of $z$
 and that the electron number density varies exponentially with $z$
 as $n_{\rm e}(z) = n_{\rm e}(0) \exp(\alpha z)$,
 where $\alpha$ is the effective ionization coefficient.
The EVDF is available from the PM calculation
 performed with only cells for velocity space.
Such a technique is to reduce the required memory capacity.

If the EVDF in a slab $z_0 \le z \le z_0 + \Delta z$ is in equilibrium,
 the electron flow at $z = z_0 + \Delta z$ is
 $\exp(\alpha \Delta z)$ times as large as that at $z$.
The PM calculation considers only $f(v, \theta)$ in the slab.
A polar-$\epsilon$ configuration was adopted
 and $\Delta z$ was set to be $\Delta \epsilon/(e E)$
 to take account of the conservation of energy \cite{Sugawara1994}.
Electrons in the slab may flow out of the slab,
 but other electrons flow into the slab from the opposite side
 to keep the electron population in the slab unchanged.
The outflow can be evaluated directly from the EVDF in the slab
 in the same way as done between parallel-plane electrodes.
The inflow is evaluated
 using the assumption of exponential spatial growth.
Let $n_{\rm f, out}$ and $n_{\rm b, out}$ be the numbers of electrons
 flowing out of the slab forward ($v_z > 0$) at $z = z_0 + \Delta t$
 and backward ($v_z < 0$) at $z = z_0$, respectively,
 and $n_{\rm f, in}$ and $n_{\rm b, in}$ be those
 flowing into the slab forward at $z = z_0$
 and backward at $z = z_0 + \Delta z$, respectively,
 during a time step $\Delta t$.
Their relations are
\begin{align}
  n_{\rm f, in} &= \exp(- \alpha \Delta z) n_{\rm f, out}, \\
  n_{\rm b, in} &= \exp(+ \alpha \Delta z) n_{\rm b, out}.
\end{align}
The value of $\alpha$ is unknown at this moment.
However, with $n_{\rm ion}$ and $n_{\rm att}$
 respectively being the electron increase and decrease
 due to ionization and attachment during $\Delta t$,
 the value of $\exp(\alpha \Delta z)$
 is obtained from a quadratic equation
 derived from the electron conservation in the slab.
They satisfy
\begin{align}
     n_{\rm f, in}
  +  n_{\rm b, in}
  -  n_{\rm f, out}
  -  n_{\rm b, out}
  +  n_{\rm ion}
  -  n_{\rm att}
 &= 0, \\
     \exp(- \alpha \Delta z) n_{\rm f, out}
  +  \exp(+ \alpha \Delta z) n_{\rm b, out}
  -  n_{\rm f, out}
  -  n_{\rm b, out}
  +  n_{\rm ion}
  -  n_{\rm att}
 &= 0.
\end{align}
This treatment is equivalent to the assumption
 $\partial/\partial z = \alpha$ and $\partial/\partial t = 0$
 in two-term approximation of the EBE analysis
 for the SST condition \cite{Thomas1969}.
$\alpha$ is obtained from
\begin{equation}
\exp(- \alpha \Delta z)
  = \frac {  n_{\rm f, out} + n_{\rm b, out}
           - n_{\rm ion}    + n_{\rm att}
 \pm \sqrt{ (n_{\rm f, out} + n_{\rm b, out}
           - n_{\rm ion}    + n_{\rm att} )^2
         - 4 n_{\rm f, out}   n_{\rm b, out} } }
          {2 n_{\rm f, out}}.
\end{equation}
The positive sign is taken in usual SST condition
 so that $\alpha = 0$ when $n_{\rm ion} - n_{\rm att} = 0$.
By a relaxation of the EVDF starting from initial EVDF and $\alpha$,
 the EVDF in equilibrium under the SST condition is obtained.

Another solution of the quadratic formula
 corresponding to the negative sign is understood to represent
 properties of electrons in backward diffusion.
In this case, $\alpha$ no longer represents
 the exponential growth of $n_{\rm e}(z)$ by ionization,
 but it represents the decay of $n_{\rm e}(z)$ toward the $-z$ direction
 in the upstream region from the electron source
 \cite{Standish1989, Sugawara1995}.
A PM calculation with such $\alpha$ \cite{Sugawara1995}
 demonstrated that the EVDF obtained in the upstream region represents
 the properties of the missing electrons \cite{Chantry1982}
 forming the decay of $n_{\rm e}(z)$ in front of the absorbing anode
 \cite{Sommerer1989, Sugawara1992}.

\subsection{Models of boundary-free real-space time-of-flight condition}

The spatial electron distribution
 $p(z, t) = \int_{\bm{v}} f(z, \bm{v}, t) {\rm d} \bm{v}$
 can be composed by superposing the $k$th-order Hermite functions
 $H_k(Z) \exp(- Z^2)$ up to a sufficient order $n$ \cite{Sugawara1997}:
 \begin{equation}
 p(z, t) = \frac{1}{\sqrt{2} \sigma(t)}
 \sum_{k = 0}^n w_k(t)
 H_k \left( \frac{z - G(t)}{\sqrt{2} \sigma(t)} \right)
 \exp \left[ - \left( \frac{z - G(t)}{\sqrt{2} \sigma(t)}
 \right)^2 \right],
 \end{equation}
 where $H_k(Z)$ is the $k$th-order Hermite polynomial
 derived sequentially from $H_0(Z) = 1$
 by a relation $H_{k + 1}(Z) = \sqrt{2} Z H_k(Z) - k H_{k - 1}(Z)$,
 $w_k(t)$ is the weight of the $k$th-order Hermite function,
 $G(t) = \langle z \rangle$ is
 the center of mass of the whole electron swarm,
 $\sigma(t)$ is the standard deviation of electron position $z$,
 and $Z = z/[\sqrt{2} \sigma(t)]$ is the dimension-less $z$ position.
$H_k(Z)$ satisfy an orthogonality
 $\int_{-\infty}^{+\infty} H_i(Z) H_j(Z) \exp(-Z^2) {\rm d} Z
 = \delta_{i j} i! \sqrt{\pi}$,
 where $\delta_{i j}$ is the Kronecker delta.
The Hermite functions are suitable to expand or compose
 Gaussian-like distributions having boundary condition
 $\lim_{z \rightarrow \pm \infty} p(z, t) = 0$,
 and it is thought that
 $p(z, t)$ of electron swarms eventually tends to a Gaussian.
Values of $w_k(t)$ are determined by the spatial moments of
 the electron swarm up to the $k$th order as shown later.

Let $m_k(\bm{v}, t)$ be
 the $k$th-order spatial moment distribution function
 with respect to the $z$ direction, and it is defined as
 \begin{equation}
 m_k(\bm{v}, t)
 = \int_{z = -\infty}^\infty z^k f(\bm{v}, z, t) {\rm d} z.
 \end{equation}
A series of moment equations in a hierarchy are derived from the EBE
 by integrating each term with weight $z^k$ over $z$
 \cite{Sugawara1997, Sugawara1998, Sugawara2017},
 and $m_k(\bm{v}, t)$ can be calculated without the partitioning of $z$:
 \begin{align}
 \frac{\partial}{\partial t} f(z, \bm{v}, t)
 &=
  -  v_z \frac{\partial}{\partial z}   f(z, \bm{v}, t)
  -  a_z \frac{\partial}{\partial v_z} f(z, \bm{v}, t)
  +  J[ f(z, \bm{v}, t) ] \\
 \frac{\partial}{\partial t} m_k(\bm{v}, t)
 &=
  +  k v_z m_{k - 1}(\bm{v}, t)
  -  a_z \frac{\partial}{\partial v_z} m_k(\bm{v}, t)
  +  J[ m_k(\bm{v}, t) ]. \label{eqn:PM-momentEq}
 \end{align}
Here, $m_0(\bm{v}, t)$ is identical to $f(\bm{v}, t)$
 representing the electron number density at $\bm{v}$.
$m_1(\bm{v}, t)$ gives
 the center of mass $G(\bm{v}, t)$ of the electrons having $\bm{v}$
 as $G(\bm{v}, t) = m_1(\bm{v}, t)/m_0(\bm{v}, t)$,
 and the center of mass $G(t)$ of the whole electron swarm is given as
 $G(t) = \int_{\bm{v}} m_1(\bm{v}, t) {\rm d} \bm{v}
       / \int_{\bm{v}} m_0(\bm{v}, t) {\rm d} \bm{v}$.
The temporal variation of $G(t)$ is the center-of-mass drift velocity
 $W_{\rm r}(t) = {\rm d}G(t)/{\rm d}t$ of the electron swarm
 \cite{Tagashira1977, Kitamori1980, Sugawara1997, Sugawara1998, Sugawara2017}.
$m_2(\bm{v}, t)$ gives the variance of the electron swarm as
  $[ \int_{\bm{v}} m_2(\bm{v}, t) {\rm d} \bm{v} ]
 / [ \int_{\bm{v}} m_0(\bm{v}, t) {\rm d} \bm{v} ] - [G(\bm{v}, t)]^2$.
Its temporal variation gives the longitudinal diffusion coefficient as
 $D_{\rm L}(t) = \frac{1}{2} ({\rm d}/{\rm d}t) \{
   \int_{\bm{v}} m_2(\bm{v}, t) {\rm d} \bm{v}
 / \int_{\bm{v}} m_0(\bm{v}, t) {\rm d} \bm{v} - [G(t)]^2 \}$
 \cite{Tagashira1977, Kitamori1980, Sugawara1998, Sugawara2017}.

For the moment calculation up to the $n$th order,
 $(n + 1)$ sets of cells are prepared,
 and initial values of $m_k(\bm{v}, t)$ ($0 \le k \le n$)
 are stored in the cells.
Their temporal variations for $\Delta t$ are calculated by applying
 the collision and acceleration propagators to $m_k(\bm{v}, t)$.
The propagators are common for $m_k(\bm{v}, t)$ irrespective of $k$.
In addition, amounts of the $k$th-order moments
 $k v_z m_{k - 1}(\bm{v}, t) \Delta t$ evaluated by the drift term
 are added to $m_k(\bm{v}, t)$ to obtain $m_k(\bm{v}, t + \Delta t)$.
The instantaneous amount of the $k$th order moment of
 the whole electron swarm is given as
 $m_k(t) = \int_{\bm{v}} m_k(\bm{v}, t) {\rm d} \bm{v}$.
$m_k(t)$, which are values in laboratory system,
 are converted to the values in center-of-mass system around $G(t)$ as
 $m'_k(t) = \sum_{i = 0}^{k} (_kC_i) [-G(t)]^{k - i} m_i(t)$.
From these quantities we can obtain higher-order
 ($n$th-order, $n \ge 3$) diffusion coefficients $D_{{\rm L}n}$ as well
 \cite{Sugawara1998};
 e.g., $D_{{\rm L}3}(t) = (1/3!) {\rm d}m'_3(t)/{\rm d}t$.
$m'_k(t)$ are further converted to to the dimension-less value $M_k(t)$
 reduced by $\sigma(t) = \sqrt{m_2(t)/m_0(t) - [G(t)]^2}$
 as $M_k(t) = m'_k(t) / [\sqrt{2} \sigma(t)]^k$.
Some first $w_k(t)$ are given as
 \begin{align}
 w_0(t)
 &=           M_0(t)
     / \left( 0! \sqrt{\pi} \right) \\
 w_1(t)
 &=           \sqrt{2} M_1(t)
     / \left( 1! \sqrt{\pi} \right) = 0 \\
 w_2(t)
 &= \left(  2 M_2(t)
            - M_0(t) \right)
     / \left( 2! \sqrt{\pi} \right) = 0 \\
 w_3(t)
 &= \left(  2 \sqrt{2} M_3(t)
          - 3 \sqrt{2} M_1(t) \right)
     / \left( 3! \sqrt{\pi} \right) \\
 w_4(t)
 &= \left(  4 M_4(t)
         - 12 M_2(t)
         +  3 M_0(t) \right)
     / \left( 4! \sqrt{\pi} \right) \\
 w_5(t)
 &= \left(  4 \sqrt{2} M_5(t)
         - 20 \sqrt{2} M_3(t)
         + 15 \sqrt{2} M_1(t) \right)
     / \left( 5! \sqrt{\pi} \right) \\
 w_6(t)
 &= \left(  8 M_6(t)
         - 60 M_4(t)
         + 90 M_2(t)
         - 15 M_0(t) \right)
     / \left( 6! \sqrt{\pi} \right) \\
 w_7(t)
 &= \left(  8 \sqrt{2} M_7(t)
         - 84 \sqrt{2} M_5(t)
        + 210 \sqrt{2} M_3(t)
        - 105 \sqrt{2} M_1(t) \right)
     / \left( 7! \sqrt{\pi} \right) \\
 w_8(t)
 &= \left(  16 M_8(t)
         - 224 M_6(t)
         + 840 M_4(t)
         - 840 M_2(t)
         + 105 M_0(t) \right)
     / \left( 8! \sqrt{\pi} \right) \\
 w_9(t)
 &= \left(  16 \sqrt{2} M_9(t)
         - 288 \sqrt{2} M_7(t)
        + 1512 \sqrt{2} M_5(t)
        - 2520 \sqrt{2} M_3(t)
         + 945 \sqrt{2} M_1(t) \right)
     / \left( 9! \sqrt{\pi} \right)
 \end{align}

The same technique to compose spatial electron distribution
 has been applied
 not only to the longitudinal direction \cite{Sugawara1998}
 but also to the transverse direction \cite{Sugawara1999}.
The higher-order transverse diffusion coefficients $D_{{\rm T}n}$
 are also available from
 the simultaneous moment equations up to the $n$th order
 with respect to the direction perpendicular to the $\bm{E}$ field
 \cite{Sugawara1999}.

\subsection{Models of velocity space and real space under uniform electric and magnetic fields}

The PM has been applied to calculation of the EVDF in equilibrium
 under dc crossed electric and magnetic fields,
 $\bm{E} \times \bm{B}$ fields,
 assuming $\bm{E} \perp \bm{B}$ as $\bm{E} = (0, 0, -E)$ ($E > 0$) and
 $\bm{B} = (0, B, 0)$ ($B > 0$) \cite{Sugawara2019b}.
The EVDF is no longer axi-symmetric
 under the $\bm{E} \times \bm{B}$ fields,
 thus it is required to have three variables to represent the EVDF\@.
A polar-$\epsilon$ configuration
 modified for three-variable velocity space $(v, \theta, \phi)$
 was chosen in the practical calculation for convenience
 in the treatment of isotropic scattering after collisions
 \cite{Sugawara2019b}.
The memory array for the EVDF becomes three-dimensional
 as the cells were prepared for every
 $\Delta \epsilon$, $\Delta \theta$, and $\Delta \phi$,
 where $v = v_1 \sqrt{\epsilon/\epsilon_{\rm 1eV}}$,
 $v_1$ is the electron speed
 associated with $\epsilon_{\rm 1eV} = 1$~eV,
 $v_x = v \sin \theta \cos \phi$,
 $v_y = v \cos \theta$, and
 $v_z = v \sin \theta \sin \phi$.
A cell may have at most six boundaries
 facing to the $\pm \epsilon$, $\pm \theta$, and $\pm \phi$ directions.
The EBE for the EVDF in velocity space becomes
 \begin{equation}
 \frac{\partial}{\partial t} f(\bm{v}, t) =
  -  \frac{e}{m} \left( \bm{E} + \bm{v} \times \bm{B} \right)
     \cdot \frac{\partial}{\partial \bm{v}} f(\bm{v}, t)
  +  J[ f(\bm{v}, t) ].
 \end{equation}
The treatment for the collision operator is unchanged
 even for the EVDF in the $\bm{E} \times \bm{B}$ fields.
On the other hand,
 the acceleration $-(e/m)(\bm{E} + \bm{v} \times \bm{B})$
 is velocity-dependent
 and electron motion in velocity space is rotational around an axis
 $(v_x, v_z) = (v_{\bm{E} \times \bm{B}}, v_{\bm{E}}) = (E/B, 0)$.
This rotation axis is parallel to the $v_z$ axis
 but off-centered (off-origin).
This arose a complexity
 in the preparation of the acceleration propagator.
The probability of the intercellular electron transition is calculated
 in the Eulerian manner
 common with that applied to two-variable velocity space,
 i.e.\ by integration of the outflow flux
 at the downstream cell boundary.
However, the cell boundaries facing to the `downstream'
 depends on the acceleration direction determined by $\bm{v}$.
Further more, the acceleration direction may change
 even in a cell boundary.
The acceleration propagator to deal with the rotational electron flow
 was prepared choosing the downstream cell boundaries carefully,
 and the order of application of the acceleration propagator
 to the cells are also arranged for fast convergence of the EVDF\@.

In an example of computation of the EVDF
 under $\bm{E} \times \bm{B}$ fields
 with $1000 \times 45 \times 720$ cells for $(\epsilon, \theta, \phi)$
 in double precision ($8$ bytes per cell) \cite{Sugawara2019b},
 the required memory capacity was roughly several GiB\@.
In addition to the array to store the numbers of electrons in the cells,
 those for associated cell properties
 such as cell volume and areas of the six cell boundaries
 are also needed.
However, the memory capacity required for this model
 is still in an available range for recent workstations.

The EBE under the $\bm{E} \times \bm{B}$ fields
 can be extended also to the spatial moment equations
 under the $\bm{E} \times \bm{B}$ fields
 with respect to the $x$, and $y$, and $z$ directions
 \cite{Sugawara2019a}
 in the same manner as Equation \eqref{eqn:PM-momentEq}
 performed in the boundary-free one-dimensional TOF model.
The center-of-mass drift velocity vector
 $\bm{W}_{\rm r} = (W_x, W_y, W_z)$
 and the direction-dependent diffusion coefficients
 $D_x$, $D_y$, and $D_z$ were calculated by the PM
 for the simultaneous spatial moment equations up to the second order.
Seven sets of cells were used to store
 the zeroth-, first-, and second-order moments
 with respect to the $x$, $y$, and $z$ directions,
 $m_{0}(\bm{v}, t)$,
 $m_{x 1}(\bm{v}, t)$, $m_{y 1}(\bm{v}, t)$, $m_{z 1}(\bm{v}, t)$,
 $m_{x 2}(\bm{v}, t)$, $m_{y 2}(\bm{v}, t)$, and $m_{z 2}(\bm{v}, t)$,
 where $m_{0}(\bm{v}, t)$ is common for the three directions.
The collision and acceleration propagators are unchanged.
Temporal development of the moments can be calculated
 by applying the propagators iteratively.
On the other hand, in case only the equilibrium values of
 $W_x$, $W_y$, $W_z$, $D_x$, $D_y$, and $D_z$ are needed,
 the relaxation of each moment can be achieved stepwise
 from the zeroth order to the second
 independently for $x$, $y$, and $z$,
 because higher-order moments depend on only the lower-order ones
 defined for the same direction via the drift term.
This feature allows us to save
 the computational load for the relaxation process
 and the required memory capacity.

\subsection{Challenges in computational techniques}

It is an advantageous feature that
 the PM can observe the temporal development of the EVDF\@.
On the other hand,
 we often need only the EVDF solution in drift equilibrium.
In conventional PM calculations,
 the equilibrium EVDF is obtained after physical relaxation of the EVDF
 under the electron acceleration by the external fields
 and scattering by gas molecules
 proceeding step by step every $\Delta t$.
This guarantees the continuity of the number of electrons.
When the difference between the normalized EVDFs derived from
 $f(\bm{v}, t + \Delta t)$ and $f(\bm{v}, t)$ is negligibly small,
 we may regard the EVDF as the converged solution.
The normalized equilibrium EVDF satisfies the EBE,
 which is a balance equation under $\partial/\partial t = 0$.
The relaxation process of the EVDF can be accelerated by a scheme
 based on the Gauss--Seidel method \cite{Sugawara2017}.
In contrast to that $f(\bm{v}, t)$ is kept unchanged
 until $f(\bm{v}, t + \Delta t)$ is obtained
 in case physical relaxation of the EVDF is observed,
 the Gauss--Seidel method renews $f(\bm{v}, t)$
 part by part (i.e.\ cell by cell)
 on the basis of the local balance expressed by the EBE
 ignoring the entire electron conservation.
It is expected here that renewed value of $f(\bm{v}, t)$ is
 closer to the equilibrium value than before renewal.
Fast propagation of the renewal result over velocity space
 is promoted by arranging the sequence of renewal calculations to be
 from the upstream cells to the downstream cells in velocity space.
This numerical relaxation process no longer has physical meaning.
However, the number of iterations
 for convergence to the equilibrium solution
 can be reduced down to the order of magnitude of $1/1000$
 in a drastic case \cite{Sugawara2017}.

A challenge to a long-term relaxation process was demonstrated
 with a PM in a matrix form \cite{Sugawara2021}.
When the EVDF at a time $t$ is represented in a vector form $\bm{f}(t)$
 with the elements corresponding to the number of electrons in each cell
 similarly to Equations
 \eqref{eq:evdf_col_vect} and \eqref{eq:n_col_vect},
 $\bm{f}(t + \Delta t)$ can be obtained by applying a propagator
 represented in a matrix form $\bm{M}$ as
 $\bm{f}(t + \Delta t) = \bm{M} \bm{f}(t)$
 in the same way as Equation \eqref{eq:time_evolution_evdf}.
Here, we may calculate powers of $\bm{M}$ as
 $\bm{P}_1 = \bm{M}^2$, $\bm{P}_2 = \bm{M}^4$,
 $\bm{P}_3 = \bm{M}^8$, $\cdots$, $\bm{P}_n = \bm{M}^{2^n}$
 by repeating the squaring operation
 for $\bm{P}_{i + 1} = \bm{P}_i \times \bm{P}_i$.
After this preparation,
 we obtain $\bm{f}(t + 2^n \Delta t) = \bm{P}_n \bm{f}(t)$.
The relaxation time $2^n \Delta t$ increases exponentially with $n$
 while computational elapse time increases linearly with $n$.
The calculation for a slow relaxation of the EVDF
 in a ramp model gas \cite{Reid1979}
 was demonstrated using $18\,000$ cells defined with
 $1000$ divisions for $\epsilon \in [0, \epsilon_{\rm max}]$
 and $18$ divisions for $\theta \in [0, \pi]$
 (thus, vector $\bm{f}(t)$ has $18\,000$ elements),
 $\Delta t = 1$~ps, and $n = 30$
 ($2^{30} > 10^9$, thus until $t > 1$~ms) \cite{Sugawara2021}.
Convergence of mean electron energy $\langle \epsilon \rangle$
 and the EVDF were observed midway by $n = 20$.
This approach enables us to observe a long-term relaxation
 in a logarithmic time scale,
 although it requires a large matrix size as $18\,000 \times 18\,000$.

Another recent improvement of the PM calculation is
 for calculations under low $E/N$ conditions and rf $\bm{E}$ fields
 \cite{Kobayashi2023}.
The polar-$\epsilon$ configuration has a tendency that
 the cells around the origin become more coarse
 than in the polar-$v$ configuration.
This feature is negligible at high $E/N$ values
 where the electron energy loss by inelastic collisions and
 the electron energy gain under high $E$ are dominant
 in the formation of EVDF satisfying the energy balance.
However, at low $E/N$ values,
 which appear not only under dc $\bm{E}$ fields
 but also in rf $\bm{E}$ fields as low $\bm{E}$ periods,
 the coarseness of the low-energy cells
 leads to an overestimation of the acceleration of low-energy electrons
 and causes a less precision in some electron transport coefficients.
This is because uniform electron distribution within a cell
 is used to be assumed in conventional PM calculations.
It was demonstrated that the overestimation is suppressed
 by blending central and upwind differences
 with relevantly chosen weights.

Owing to the progress in both algorithm and computational facility,
 the PM is nowadays recognized
 as an available and possible self-consistent EVDF solver.
With already established techniques,
 the PM can be used not only to solve
 the EVDF in typical observation models
 but also for preparation of look-up tables of
 a set of electron transport coefficients,
 mean electron energy $\langle \epsilon \rangle$,
 ionization coefficient $\nu_{\rm ion}$,
 drift velocity $W$, and diffusion coefficient $D$
 under uniform dc $\bm{E}$, for the use in fluid simulations.
Further enhancement of the calculation speed and memory capacity
 would arrow us consideration for
 more practical effects of electron--molecule interactions,
 higher-order multi-dimensional cell configurations,
 and more complicated geometries of the plasma reactors.
Some of the basic processes not dealt with sufficiently would be
 anisotropic scattering, super-elastic collisions,
 Coulomb collisions between electrons, etc.
For the cell configuration, realization of PM calculation
 in dimensions higher than three
 would have to wait for more empowerment of the computers;
 e.g., self-consistent PM calculation in a cylindrical symmetry
 requires five-variable phase space as $(r, z, v, \theta_v, \phi)$,
 which requires a tremendously huge memory capacity
 even with a coarse resolution for each variable.
Nonetheless, for frequently assumed geometries
 such as one-dimensional real space between parallel-plane electrodes,
 the PM calculations done for three-variable phase space
 would withstand requirements on practical conditions
 by its flexibility for linkage with,
 for example, Poisson's equation and boundary conditions.

\section{Summary and conclusions}

The methods that have been described in this review have in common that they describe the kinetics of electrons, subject to the effect of electric and magnetic fields and collisions with atoms and molecules in a low-temperature plasma, using a propagator-based description. Propagators are operators that shuffle the probability distribution of the electronic fluid in the phase space, moving it from one position to another in this generalized space: as many mathematical objects of a certain importance can have a description apparently very different from another depending on of the formal language on which the initial setting of the algorithm is based. This means that in some cases, such as in the writing of a so-called traditional MC program \cite{longo2000monte}, the role of propagators may be much less evident at a superficial level. In some other cases \cite{Sugawara1992}, propagators enter not only into the settings of the algorithm but right from the beginning of the language that is used to describe it. A hybrid stochastic-deterministic approach, such as the MCF \cite{schaefer1990monte, vialetto2019benchmark}, which effectively uses the propagators twice, first as ``micro-propagators'' for calculations of transition probabilities, and then as Markov matrices to extend the simulation to the characteristic times of energy diffusion, is an even more particular case. 

Emphasizing the common roots of these approaches and others that are comparable, does not erase the fact that in the concrete study of a certain type of ionized gas rather than another a method in particular may be much superior to others: elements that contribute to this choice are the relationships between the various characteristic times of the plasma, the type of collision processes, the dimensionality of the system, the presence and complexity of boundary conditions, even practical requirements such as the reduction of requested memory, speed and computer time, the physical transparency of the method and ease of code maintenance. However, the ability to understand the conceptual bases and common aspects from a mathematical point of view of the calculation methods available for the study of the electron kinetics of low-temperature plasmas can represent a very useful element of knowledge, both for researchers who have to decide the best strategy for the representation of a system and for instructors who teach these methods and arouse interest in them. 

In all cases, the fundamental object in question is a time-dependent relationship between two states in phase space and which can be described physically as a function or as an operator acting on a function, and numerically as a matrix. Hence, its treatment requires tools from algebra and calculus. 
%This provides highly technological, useful and monetizable applications for calculus and algebra concepts often considered abstract. 
It is possible that the current approach to the simulation of plasma-based application systems, which is a modular approach, may hide the conceptual interest of these calculation tools: they constitute the most physical part of simulation programs and their connection with other areas of physics and mathematics are both aspects that can spark new interest in these tools. The mathematics of propagators allows us to solve the problem of the kinetics of plasma electrons with great efficiency, and this makes these instruments useful tools, because the kinetics of electrons in plasma cannot be neglected in reliably calculating chemical reaction rates and transport quantities. 

Understanding the mathematical concept underlying a group of models is a stimulating exercise in itself and also has a heuristic value, because it produces future research questions: 
\begin{itemize}
  \item What is the most efficient presentation of the propagation concept depending on the conditions of plasma?
  \item How can the concept of propagation be extended to nonlinear systems such as those that emerge when collisions between charged particles or gas heating are taken into account?
  \item Can the mathematical grids that are employed at different times of the different approaches be replaced with spectral descriptions?
  \item  Could propagators, which represents input-output relations, efficiently be computed using machine learning and artificial intelligence?
  \item Could the similarity between the propagators used in plasma and those used in other areas of physics help to develop faster or less memory-intensive computational methods?
\end{itemize}

More recent applications of low-temperature ionized gases from biology to space science to materials science will always produce new possibilities for numerical descriptions of electron transport that have new strengths even as they have their specific weaknesses. The authors hope that this review work can stimulate research activity that captures new insights and develops new languages. 

\section*{Acknowledgements}
LV acknowledges financial support by the Stanford Energy Postdoctoral Fellowship through contributions from the Precourt Institute for Energy, Bits \& Watts Initiative, StorageX Initiative, and TomKat Center for Sustainable Energy.

\section*{References}
\bibliographystyle{myieeetr}
\bibliography{Paperbib}

\end{document}